\documentclass[10pt,letterpaper,twocolumn]{article}%% two column, final layout

\usepackage{ol2}
\usepackage[draft]{hyperref}
\usepackage{amsmath}

\begin{document}

\twocolumn[ %% activate for two-column option

\title{Mode Locking At and Below the CW Threshold}

%% For REVTeX it is possible to automate superscript and e-mail callouts with the superscriptaddress option; see REVTeX4 documentation.

\author{Shai Yefet and Avi Pe'er$^{*}$}

\address{
Department of physics and BINA Center of nano-technology, Bar-Ilan university, Ramat-Gan 52900, Israel

$^*$e-mail: avi.peer@biu.ac.il
}

\begin{abstract}
We explore experimentally a new regime of operation for mode locking in a Ti:Sapphire laser with
enhanced Kerr nonlinearity, where the threshold for pulsed operation is lowered below the threshold
for continuous-wave (CW) operation. Even though a CW solution cannot exist in this regime, pulsed
oscillation can be realized directly from zero CW oscillation. In this regime, the point of maximum strength of the Kerr nonlinear
process provides a "sweet spot" for mode locking, which can be optimized to considerably lower the pump power threshold. The properties of the
"sweet spot" are explained with a qualitative model.
\end{abstract}

%140.3538   Lasers, pulsed
%140.3580   Lasers, solid-state
%140.7090   Ultrafast lasers
 ] %% activate for two-column option

\noindent The ultra-broad gain bandwidth of the Ti:Sapphire (TiS) laser renders it the 'work-horse' of the last decades for generation
of ultrashort pulses by mode locking (ML) \cite{Haus@modelocking}. The nonlinear mechanism responsible for ML is self-focusing of the beam due to the optical Kerr effect within the TiS crystal, introducing an intensity dependent loss mechanism that favors
pulses over continuous-wave (CW) operation \cite{hardaperture}. A known feature of ML is the abrupt transition between CW and ML operation in terms of pump power \cite{Ohno@abruptmodelocking1978}.
%\cite{Ohno@abruptmodelocking1976,Ohno@abruptmodelocking1978}.
Only when the pump power crosses a certain threshold, ML can be initiated from
a noise-seeded fluctuation (either by a knock on a cavity element or by external injection of long pulses).
Another common feature is that the threshold pump power for ML is
higher than the CW threshold. It seems as if a certain amount of CW oscillations is necessary, and only on top of an existing CW can an intensity fluctuation be amplified to create the pulse.

The threshold-like behavior of ML was elegantly explained by the theory
of statistical light-mode dynamics (SLD) \cite{SLD@theory2},
%\cite{SLD@theory1,SLD@theory2,SLD@theory3}
where the transition from CW to ML is described as a first order phase transition,
in which the order parameter (analogous to temperature) is $T\sim1/(\gamma_{s}P^{2})$, where $\gamma_{s}$
represents the strength of the relevant nonlinearity and $P$ is the total laser cavity power.
It was demonstrated \cite{SLD@exp2}
%\cite{SLD@exp1,SLD@exp2}
that on top of an existing CW, ML can occur only when $T$ is lowered below a critical value of $T_{c}$. Yet, to our knowledge, the question whether the
preliminary existence of a CW oscillation is a necessary condition for ML operation was not directly
explored and is not trivial to answer a priori. Here we demonstrate experimentally that an initial CW power need not exist.

The addition of a second intracavity Kerr medium was
explored in the past using different combinations
of gain and Kerr media \cite{n2additional1,n2additional2,n2additional3,n2additional4,n2additional5,n2additional6},
and was shown to lower the ML threshold.
Here, we further enhance the nonlinear Kerr mechanism, observing a new regime of mode locking,
where: 1. the intracavity CW power needed to initiate ML can be reduced to zero, 2. pulses can be sustained even below the
CW threshold and the pump power necessary for pulsed operation can be considerably improved.

\begin{figure}
\centerline{\includegraphics[width=7.5cm,trim=0cm 1cm 0cm 1cm]{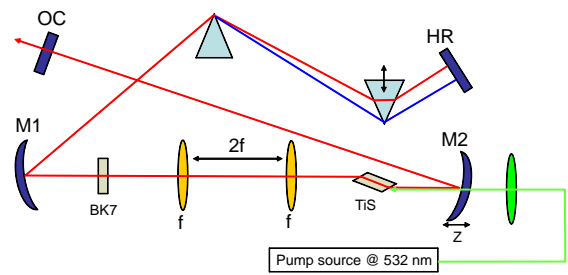}}
\caption{\label{cavityTisapSF6} Cavity configuration. The gain medium is a $3mm$ long Brewster-cut
TiS crystal with $0.25$ wt$\%$ doping. The curved mirrors ($M1\!,\!M2$) radius of curvature is
$R\!=\!15cm$, with high reflector (HR) and a $95\%$ output coupler (OC) as end mirrors. An additional
planar-cut BK7 window is inserted near the image point
of the TiS crystal, created by the two-lens telescope of focal length $f\!=\!10cm$. The short cavity arm is $42cm$ long and the long arm ($90cm$) contains a prism-pair of BK7 glass ($60cm$). Each cavity mirror except the OC provides $GDD\!\approx\!-55fs^{2}$.}
\end{figure}

Our linear TiS cavity is illustrated in Fig.\ref{cavityTisapSF6}. By adding a lens based $1$x$1$ telescope between the curved mirrors the focus inside the TiS crystal is imaged towards mirror $M1$, allowing us to enhance the nonlinearity of the cavity in a controlled manner by introducing an additional
Kerr medium near the imaged focus while varying its position.
We first introduced a $3mm$ long planar window of BK7 glass, which was AR coated and set near normal incidence. As opposed to Brewster windows, where the beam expands in one dimension due to refraction, thereby reducing the nonlinear response and generating an astigmatic Kerr lens, with normal incidence the intra-cavity beam retains its small size, which enhances the nonlinearity and provides an astigmatic-free Kerr lens.

If we define $\delta$ as a measure for the distance between $M1$ and $M2$ with respect to
an arbitrary reference point, two separate bands of $\delta$ values $[\delta_{1},\delta_{2}],\ [\delta_{3},\delta_{4}]$
allow stable CW operation of the cavity. These two stability zones are bounded by four stability
limits ($\delta_{4}>\delta_{3}>\delta_{2}>\delta_{1}$) and the working point for ML in our experiment
is near the second stability limit $\delta_{2}$ \cite{hardaperture}. Near this limit the additional nonlinear Kerr lens
causes a decrease of the mode size at the OC for ML. In order to favor ML, one can
exploit the Kerr lens in one of two ways: either by placing an aperture near the OC to selectively
induce loss on the CW mode, or by increasing the distance $\delta$ between the
curved mirrors a little bit beyond the stability limit of $\delta_{2}$, which passively
induces diffraction losses to the CW mode. For pulsed operation, the additional Kerr lens re-stabilizes the cavity, eliminating the diffraction losses. In this manner increasing the distance $\delta$ is equivalent to closing a physical aperture on the beam, which increases the threshold for CW due to loss, and requires higher power in ML for the nonlinear lens to overcome the loss.

\begin{figure}[htb]
\centerline{\includegraphics[width=8.5cm,trim=0cm 0.5cm 0cm 0cm]{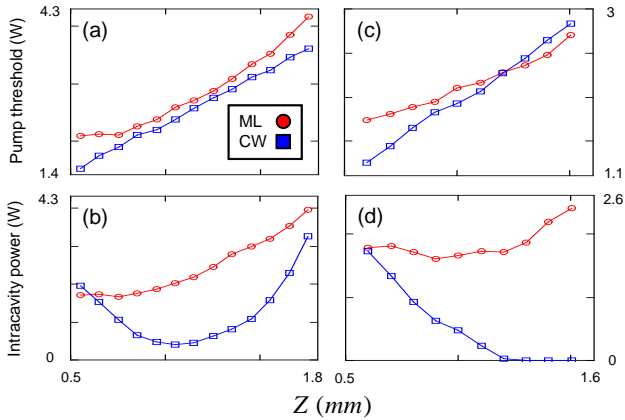}}
\caption{\label{CWMLparameters} ML (red) and CW (blue) operation parameters as a function
of $Z=\delta-\delta_{2}$ for two $3mm$ long BK7 window positions: off-focus (a)+(b) and in-focus (c)+(d).}
\end{figure}

The measured ML and CW operation parameters are plotted in Fig.\ref{CWMLparameters} as a function of $Z=\delta-\delta_{2}$.
Measurements were taken for two positions of the BK7 window: 1. at the imaged focus, where considerable nonlinearity is added by the window (in-focus). 2. several centimeters away, much beyond the Rayleigh range of the intracavity mode (off-focus), where the added nonlinearity is negligible and the cavity acts as a standard TiS cavity (with some additional material dispersion). Fig.\ref{CWMLparameters}(a) plots the CW
threshold and the ML threshold as a function of $Z$ for off-focus position. The ML threshold
is defined as the minimum pump power required to initiate pulsed operation. As expected, the CW threshold increases with $Z$,due to increased diffraction losses. The ML threshold also increases (since higher power is needed for ML to overcome the loss), yet with a varying slope as $Z$ increases. Fig.\ref{CWMLparameters}(b) plots the CW and ML intracavity powers at the ML threshold as a function of $Z$ for off-focus position.
As  typical for ML lasers, the ML threshold is always larger than
the CW threshold and CW oscillation must exist to initiate the ML process. In addition, although ML
operation is favorable over the entire range of $Z$, it is most favorable at the
"sweet spot" ($Z_{ss}\approx1.1mm$) where the CW oscillation required to start the
ML process reaches a minimal value.

The same CW and ML parameters are plotted in Fig.\ref{CWMLparameters}(c) and (d)
for in-focus position. The ML threshold (Fig.\ref{CWMLparameters}(c)) is reduced by
the added nonlinearity, and the ML threshold curve eventually crosses the CW
threshold at $Z_{c}\approx1.2mm$ where the intracavity CW power drops to zero,
marking the transition point to a different regime. At $Z_{c}$, ML can be achieved
from pure fluorescence with no CW oscillation. The corresponding CW and ML
intracavity powers at the ML threshold are shown in Fig.\ref{CWMLparameters}(d).
Beyond the crossing point ($Z>Z_{c}$), stable ML can still be initiated, but only by
first raising the pump power up to the CW threshold, locking, and then lowering the pump
again. At the CW threshold the pump power is too high and mode locking generates a pulse with a CW spike
attached to it, which can be eliminated by lowering the pump power below the CW threshold.
The ML threshold in Fig.\ref{CWMLparameters}(c)
for $Z>Z_{c}$ is the minimal pump power needed to maintain a clean pulse.

We can understand the need to first increase the pump power to the CW threshold and than lower it by noting
that the CW threshold marks the crossover between decay and amplification
in the cavity. For ML to occur, an intensity fluctuation must first be linearly amplified to
a sufficient peak power to initiate the Kerr-lensing mechanism. For $Z>Z_{c}$, one must pump the laser sufficiently for a noise-induced
fluctuation to be amplified (rather than decay) in order for it to reach the peak intensity
required to mobilize the Kerr-lensing process. After reducing the pump power to the
ML threshold, a clean pulse operation is obtained, but if ML is broken the cavity will not mode-lock again.

\begin{figure}[htb]
\centerline{\includegraphics[width=8.8cm,trim=0cm 1cm 0cm 0cm
]{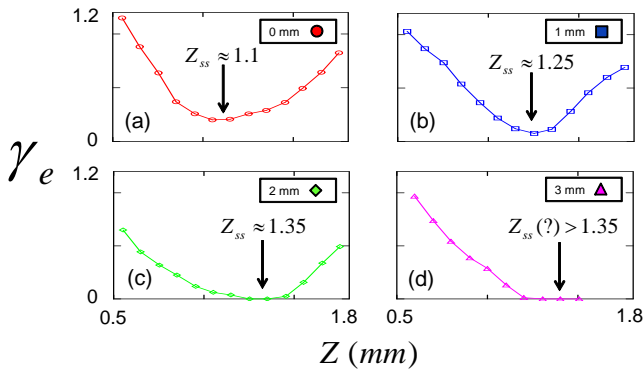}}
\caption{\label{BK7lengths} Experimental definition of the Kerr strength as a function of $Z$ for BK7 window with different lengths.}
\end{figure}

To investigate the appearance of the new regime ($Z>Z_{c}$) for in-focus window position, we plot the ratio of CW to ML powers $\gamma_{e}\!\equiv\!P_{CW}\!/\!P_{ML}$ as a function of $Z$, for windows of variable thickness (Fig.\ref{BK7lengths}). $\gamma_{e}$ represents an experimental measure for the strength of the Kerr effect, which demonstrates a "sweet spot" where $\gamma_{e}$ is minimal and the nonlinear mechanism is most efficient. The apparent tendency from Fig.\ref{BK7lengths} is that for increased nonlinearity, the sweet spot is pushed to larger $Z$ and the $\gamma_e$ value at the sweet spot is reduced. We find that for a $2mm$ thick window the $\gamma_e$ curve touches on zero near the sweet spot, marking the onset of the new regime. For a $3mm$ thick window the curve crosses zero at $Z\!=\!Z_{c}$. Well above $Z\!>\!Z_c$ pulsed operation becomes unstable, and we could not observe the reappearance of the $\gamma_e$ curve for larger values of $Z$. At every experimental point the prisms were adjusted to provide the broadest pulse bandwidth. The maximum bandwidth was obtained near the sweet spot due to the maximized Kerr strength, reaching $\approx\!100nm$ for all of the window thicknesses. This indicates that the bandwidth was limited mainly by high order dispersion of the prisms-mirrors combination, and not by the added dispersion of the windows. Although the pulse temporal width was not measured, we expect the pulses to be nearly transform limited based on previous experience with similar TiS cavities.

To provide a qualitative model for the dynamics of the "sweet spot" with increasing Kerr nonlinearity we examine a commonly used theoretical measure for the Kerr Strength:
\begin{equation}
\gamma_{s}\equiv\frac{P_{c}}{\omega}\frac{d\omega}{dP} , \label{gamma}
\end{equation}
where $\omega$ is the mode radius at the OC and $P$ is the pulse peak power normalized to the critical power
for self-focusing $P_{c}$ \cite{criticalpower}. This Kerr strength, which represents the change of the mode size due to a small increase in the ML power $P$, is a convenient measure for mode-locking with an aperture near the OC. Yet, since increasing $Z$ beyond $\delta_2$ is equivalent to closing an aperture, $\gamma_{s}$ is useful also for our configuration. Usually, $\gamma_{s}$ is calculated at zero power ($P=0$) \cite{gammadefinition} to estimate the tendency of small fluctuations to develop into pulses. We note however that the dependence of $\gamma_{s}(P)$
on power is most important. Specifically, a large (negative) value for $\gamma_{s}$ indicates that only a small increase in the ML power (or threshold) will be necessary to overcome a small reduction of the aperture size (or increase in $Z$). The power where $\gamma_s$ is most negative will represent a sweet spot for mode locking.

\begin{figure}[htb]
\centerline{\includegraphics[width=8.5cm,trim=0cm 1cm 0cm 0cm
]{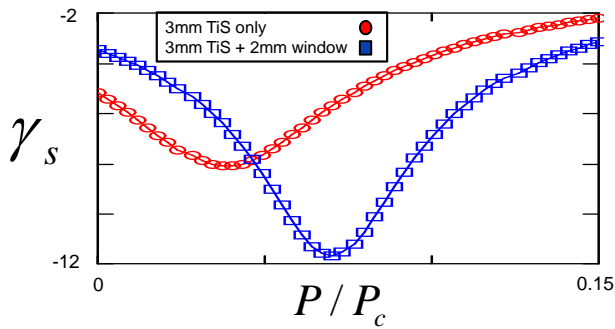}}
\caption{\label{model} Theoretical definition of the Kerr strength as a function of $P$ for TiS crystal with added window.}
\end{figure}

Figure \ref{model} plots $\gamma_{s}(P)$ for no added Kerr window (TiS only) and for a $2mm$ long added window, demonstrating a clear minimum (sweet spot) on both curves. Furthermore, as Kerr material is added (enhanced Kerr strength), the minimum point is deepened and pushed towards higher power $P$ (larger $Z$), similar to the observed $\gamma_{e}$. Although $\gamma_s$ and $\gamma_e$ are somewhat different measures for the Kerr strength, the calculation of $\gamma_s$ provides reasoning for the measured behavior of the sweet spot for the different window thicknesses.

In conclusion we note that the performance of the ML laser reported here at the critical distance $Z_c$, where the thresholds for pulsed and CW operation meet in Fig.\ref{CWMLparameters}(c), can be compared to recently published record-results \cite{lowpowerpumpTiS} that reported a mode-locked TiS laser with low pump power of $2.4W$ using an OC of 99\% and curved mirrors radii of $R=8.6cm$ with output power of $30mW$ and intracavity power of $3W$. Here, we have achieved ML from zero CW oscillation with similar repetition rate ($\approx80MHz$) using pump power of $2.3W$ with far less stringent conditions. In our experiment, the OC had only 95\% reflectivity (5 times more losses), coupling more power out ($\approx85mW$) with lower intracavity power of $1.7W$ and using curved mirrors of radius $R=15cm$. With further optimization of the added window material and cavity parameters, this may allow the development of ultra-low threshold ML sources in the future.

This research was supported by the Israeli science foundation (grant 807/09).

%\bibliographystyle{osajnl}
%\bibliography{shai}

\end{document}